\documentclass[journal]{vgtc}   
\ifpdf
  \pdfoutput=1\relax                   
  \usepackage{graphicx}                
  \DeclareGraphicsExtensions{.pdf,.png,.jpg,.jpeg} 
\else
  \usepackage{graphicx}                
  \DeclareGraphicsExtensions{.eps}     
\fi%

\graphicspath{{figures/}{pictures/}{images/}{./}} 

\usepackage{microtype}                 
\PassOptionsToPackage{warn}{textcomp}  
\usepackage{textcomp}                  
\usepackage{mathptmx}                  
\usepackage{times}                     
\usepackage{cite}                      
\usepackage{tabu}                      
\usepackage{booktabs}                  
\usepackage{balance}       
\usepackage{graphics}      
\usepackage[T1]{fontenc}   
\usepackage{txfonts}
\usepackage{mathptmx}
\usepackage[pdflang={en-US},pdftex]{hyperref}
\usepackage{color}
\usepackage{booktabs}
\usepackage{textcomp}

\usepackage{microtype}        
\usepackage{ccicons}          

\usepackage{todonotes}

\usepackage{soul}
\usepackage{balance}
\usepackage{enumitem}

\usepackage{multicol}
\usepackage{float}
\usepackage{color, colortbl}
\usepackage{csquotes} 

\definecolor{lgray}{gray}{0.9}
\definecolor{orange}{rgb}{1,0.5,0}
\graphicspath{ {figs/} }

\onlineid{0}
\vgtccategory{Research}
\vgtcpapertype{theory/model}
\title{Communicative Visualizations as a Learning Problem}
\author{Eytan Adar and Elsie Lee}
\authorfooter{
\item
 Eytan Adar and Elsie Lee are with the University of Michigan. E-mail: {\{eadar, elsielee\}@umich.edu.}
}

\shortauthortitle{Adar and Lee: Communicative Visualizations as a Learning Problem}

\abstract{
Significant research has provided robust task and evaluation languages for the analysis of \textit{exploratory} visualizations. Unfortunately, these taxonomies fail when applied to \textit{communicative} visualizations. Instead, designers often resort to evaluating communicative visualizations from the cognitive efficiency perspective: ``can the recipient accurately decode my message/insight?'' However, designers are unlikely to be satisfied if the message went `in one ear and out the other.' The consequence of this inconsistency is that it is difficult to design or select between competing options in a principled way. The problem we address is the fundamental mismatch between how designers want to describe their intent, and the language they have.  We argue that visualization designers can address this limitation through a learning lens: that the recipient is a student and the designer a teacher. By using learning objectives, designers can better define, assess, and compare communicative visualizations. We illustrate how the learning-based approach provides a framework for understanding a wide array of communicative goals.  To understand how the framework can be applied (and its limitations), we surveyed and interviewed members of the Data Visualization Society using their own visualizations as a probe. Through this study we identified the broad range of objectives in communicative visualizations and the prevalence of certain objective types.
}

\teaser{
  \centering
  \includegraphics[width=0.9\textwidth]{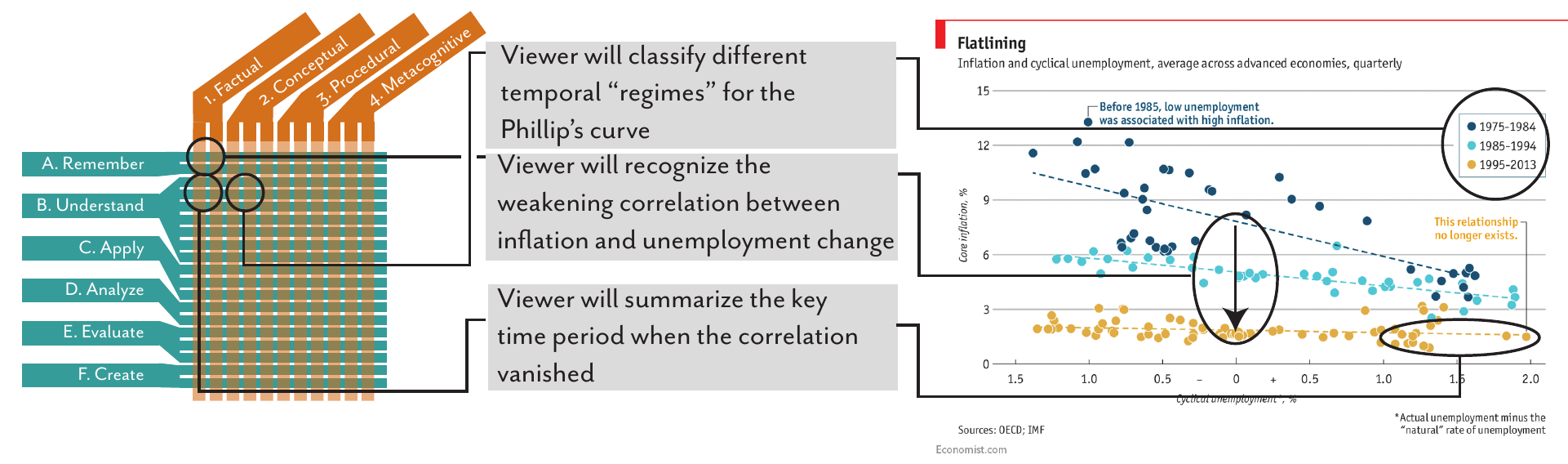}
  \caption{The extended Bloom Taxonomy (left) and a visualization of the Phillips curve (right)~\protect\cite{econcurve}. Learning objectives, and visual support for them, link the two images.}
\label{fig:phillips}
}

\vgtcinsertpkg

\begin{document}



\maketitle
\section{Introduction}
\textit{Communicative} visualizations represent the bulk of exposure any individual has to visualizations. We experience the messages of data journalists, scientists, instructors, designers, and analysts as charts, graphs, and in many other forms. In each case, the person creating the visualization or context (the thing---a paper, article, etc.---in which the visualization was embedded) has a specific set of intents. The intents are as unique as the visualizations with which they are associated: A journalist may seek to explain an insight; a scientist or analyst to convey evidence or to support a decision; an instructor to teach the relationship between two interacting chemicals. The main question we tackle here is: \textit{how do we formally describe communicative intent in visualizations?} We propose that using \textit{cognitive learning objectives} as a frame will encourage a better way of building communicative visualizations. 

With apologies to Bloom~\cite{bloom1956taxonomy}, learning objectives may help address our problem because they are, ``explicit formulations of the ways in which [\textit{viewers} (i.e., students)] are expected to be \textit{changed} by [\textit{communicative visualizations} (i.e., the educative process)].'' In their role as `educational tools', communicative visualizations must be designed as ``\textit{intentional} and \textit{reasoned} act[s]''~\cite{anderson2001taxonomy}. Doing so requires a formal language to allow a designer to explicitly formulate their expectations and intents. 

Given the prevalence of advice and taxonomies for visualization designers, it is worth asking why we even need such an `intent language?' Significant literature already exists to ensure that our viewer can \textit{read} our encoding of data accurately and effectively---a success, if that was really the designer's intent. However, knowing that the visualization will support finding $X$, or the encoding will allow the viewer to accurately decode $Y$, is poor proxy for knowing if the visualization satisfied our communicative intent. A designer would not, and should not, be satisfied if the message was, `in one ear and out the other.' Knowing the message was communicated clearly and interpreted accurately may be necessary, but is not sufficient.

\begin{figure}[htbp!]
  \centering
\includegraphics[width=\linewidth]{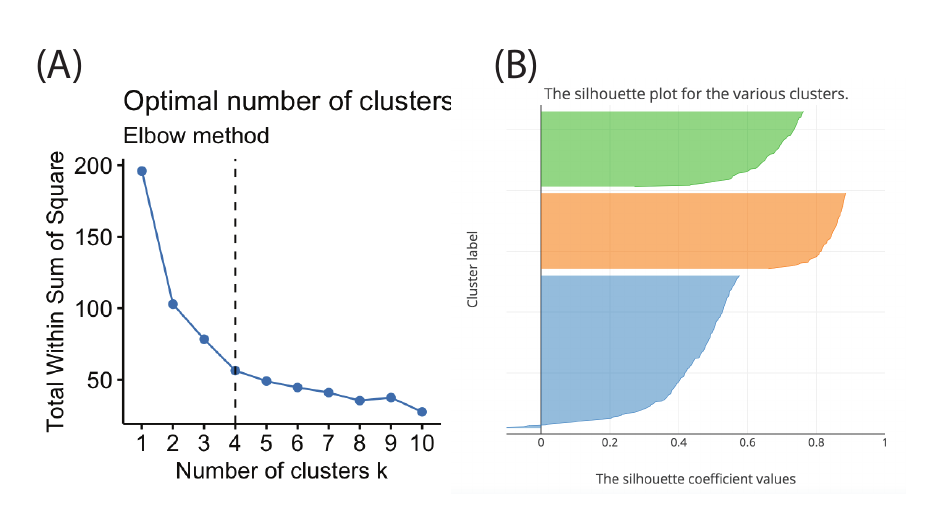}
  \caption{Two visualizations depicting performance for a clustering algorithm based on the number of clusters (cluster count is a tunable parameter): (A) the elbow method~\protect\cite{kmeans} and (B) silhouette plot~\protect\cite{sil}.}
\label{fig:elbow}
\end{figure}

Existing task and evaluation taxonomies are not refined enough to describe the intent behind a communicative visualization. Take as a simple example the plot in Figure~\ref{fig:elbow} which we may encounter reading a technical paper, webpage, or textbook. The plot shows the Sum of Squares Distances between entities as a function of the number of clusters. It is used in k-means clustering for the `elbow method' of determining an optimal $k$~\cite{kmeans} (roughly, that one should pick the number of clusters where there is a `kink' in the plot, e.g., 4 clusters). The plot in our context is communicative--it was produced by someone else to tell us, the readers, \textit{`something'}. That `something' reflects the designer's many possible intents. This may be to convince us that a choice of $k=4$ was correct; to relay the insight that $4$ was significantly better than $k=3$ or $k=5$; to critique a bad choice of $k$; to teach us what the term `elbow' means; to demonstrate how to read or create a plot suitable for an elbow method analysis; to contrast it to an alternative (e.g., the silhouette plot, Fig.~\ref{fig:elbow}B); or to lead us to create alternatives.

All these are possible--in fact, likely--intents. But how does the designer know that the visualization is successful? The mechanisms for evaluating are as varied as the intents: Can the viewer recall which $k$ was picked? Can they define an elbow point? Can they read a new plot? Can they produce a similar plot for their own data? Can they critique different plots? Offer new ones? Current taxonomies of visualization are insufficient both to describe and evaluate communicative intent. While many task/evaluation taxonomies are deeply detailed in regards to `analytical' tasks, they lump together all communicative tasks into one category. To build and evaluate communicative visualizations requires \textit{a refined and principled language for describing communicative intent}.

We argue that a good language for describing intents is the language of \textit{learning objectives}. Learning objectives allow an instructor, in this case, the designer, to explicitly define how they want to impact the viewer. For example, the designer may say, ``After seeing the visualization the viewer will be able to define an elbow point'' or ``the viewer will be able to categorize different types of students based on performance.'' None of this is to say that the viewer's needs, goals or high-level tasks should be ignored. An effective designer will integrate the viewer's objectives into the defined learning objectives--making them part of the intent. The advantage of this formulation is we can describe our objectives whether or not we know what the viewer wants (or when the viewers themselves can't describe their need). More critically, a specific learning objective is assessable. That is, the designer can craft specific tests that validate if the  visualization leads to viewers achieving the objective. Designers can compare different visualization choices using objectives and assessments. The added advantage is that we can ask if our objectives are achieved not only when the viewer is \textit{looking at the visualization} but, more importantly, \textit{when it was taken away}. This is different from most evaluative techniques that focus on what happens when the visualization is in front of the viewer.

We offer a taxonomy to satisfy our goal of modeling designer intents. This emerges from the original learning objectives literature~\cite{bloom1956taxonomy}--progenitor of most modern learning objective languages. The taxonomy allows a designer to construct statements of the form: \textbf{The viewer will \textit{[verb]} \textit{[noun]}}.\footnote{We regard the form \textit{viewer will recall \ldots} and \textit{will be able to recall \ldots} and \textit{will learn to recall \ldots} as equivalent. However, we prefer the first for conciseness and to avoid nominalization.} This \textit{cognitive} taxonomy (Figure~\ref{fig:cogtax}) will provide classified verbs related to cognitive processes (e.g., recall, explain, critique, etc.) and nouns related to types of knowledge (e.g., insights, algorithms, etc.). The formalism and structure of this language will allow us to create assessments to evaluate visualizations. To validate our taxonomy we surveyed (n=29) and interviewed (n=16) visualization design professionals to identify if, and how, the taxonomy could be used to specify their design intents. The complete taxonomy and example visualizations are available at \url{http://visualobjectives.net}. 

In the interest of practicing what we preach, we offer our contributions in the form of learning objectives. Upon reading this article, the reader will be able to \ldots 

\begin{itemize}[noitemsep]
    \item \ldots identify how communicative visualization is like a teacher/student relationship.
    \item \ldots describe the limitations of existing task taxonomies for communicative visualizations.
    \item \ldots apply the objectives framework to describe communicative visualization intents.
    \item \ldots identify how designers map the objectives taxonomy to intents with real examples.
    \item \ldots apply the framework to their own visualizations.
\end{itemize}

\section{Related Work}
A motivation for creating a new taxonomy is the limitation in existing forms for communicative intent. Ideally, our alternative should: (1) map to the language a designer might use, (2) cover the breadth and depth of communicative intent, (3) work at appropriate granularity, and (4) lead to appropriate and convincing assessments of the visualization.

Before describing related work, we offer one point on notation. A key feature of most existing taxonomies is their focus on the single `agent' at work. This is because most exploratory visualization systems have one agent to consider: the `analyst.' Analysts take raw data, and driven by some motivating questions, will use the visualization to find an answer (pattern-finding) or identify structures in the data (pattern-making).  In contrast, with communicative visualizations, we have at least two agents: the sender and the recipient. In a formal learning environment, the sender might be a teacher and the recipient, the student. A visual journalist may send a message to the reader. The scientist may send the message to their peers. 

Unlike the singular analyst, the senders and recipients may have different reasons for using the visualization as a communicative medium. The sender \textit{designs} the \textit{context for transmission} (e.g.,  an article, a web page, a scientific paper, or an argument) that contained the visualization. This may or may not include designing the visualization itself (i.e., the \textit{sender} may not be the \textit{creator}). In each case--teacher, journalist, scientist--the sender may not be the author of the visualization but an intermediary. To convey this range, we adopt \textit{designer} and \textit{viewer}. 

A distinguishing feature of this model is that the viewer's access to the information is mediated by the designer's choices. Whatever evidence the designer provided or however they distilled the data will impact what the viewer can do. This observation is a critical feature of communicative visualizations: a human agent has acted in shaping the message, and in doing so, the designer can shape the viewer.

A user-centered approach to visualization design (i.e., viewer-centered) can not account for situations in which the viewer's needs are different from the designer's. In such cases, the designer's intents take precedence over the viewer's. A benefit of the learning objectives framework is that it accounts for this potential power differential.

\subsection{Task Taxonomies}
Taxonomies of visualization tasks often put communication as a broad category. In some cases, communicative visualizations are placed in the context of analytic tasks such as ``learning about data''~\cite{Fekete2008} or ``domain''~\cite{Amar:2005:LCA:1106328.1106582,6634168}. Often communication is embedded in a workflow an analyst would use: first I find the insight \textit{and then} I present it. This focus is most apparent in the context of collaborative information visualization where the analyze/communicate cycle is constant~\cite{heer2007voyagers, viegas2007manyeyes, nobarany2012facilitating, 5386648}. Consequently, many taxonomies do not distinguish between different communicative goals. They relegate communicative visualization into a single, simple abstract box. But this box is complex--it must reflect the intent of the designer, the goals and needs of the recipient, the communication context, and the interaction between all these factors. Unpacking this box can enable better design and evaluation strategies.

The other areas of research and practice that have \textit{much} to say about communicative visualization are those applying the lens of cognitive efficiency--mechanisms to ensure that the viewer can accurately decode the message. There is no shortage of advice to designers, and many case studies and research results from psychology and cognitive sciences have been distilled into numerous books (e.g., \cite{evergreen2016effective, few2004show, cairo2012functional,  knaflic2015storytelling, tufte2001visual, wong2013wall}). All of this is useful if the designer needs to know what type of chart best supports reading a correlation statistic but offers less when it comes to other cognitive tasks and evaluation. Generalized systems such as APT, Tableau, and Grammar of Graphics based tools~\cite{mackinlay1986automating,729564,6634156,wilkinson2006grammar} leveraged this research to recommend the best visualization given the `data of interest' or broad analytical targets. However, while guidelines lead to better visualizations, they do not allow the end-user (in our case, designer) to explicitly specify their objectives or insights--broadly, their communicative intent. The consequence is that the guidelines and tools can't help to assess the produced images. By incorporating the learning objectives frame, such tools could be enhanced for communicative visualization tasks.

\subsection{Evaluation Taxonomies}
In addition to \textit{task} taxonomies we can consider the many \textit{evaluation} taxonomies and specific techniques for different contribution types~\cite{4035759,lam:hal-00723057}. These techniques range in focus from perception~\cite{Zuk:2006:HIV:1168149.1168162}, to usability~\cite{Shneiderman:1996:ETD:832277.834354}, and to the discovery process~\cite{amarstasko}. Given the focus on visual analytics tools, evaluation research often focuses on sense-making and insight~\cite{Russell:1993:CSS:169059.169209}, with a particular challenge in demonstrating ecological validity~\cite{Saraiya:2005:IME:1070610.1070747, North:2006:TMV:1137231.1137267, Yi:2008:UCI:1377966.1377971,4797511}. Believable studies are often longitudinal~\cite{Shneiderman:2006:SEI:1168149.1168158} or use high-cost observation approaches ~\cite{Isenberg:2008:GEI:1377966.1377974}. Unfortunately, these techniques are not well suited for evaluating communicative visualizations, which are unique and numerous.  Visual analytics processes are `pull' driven where the user introduces many of the constraints and demands. With communicative visualization the idea of the user is vaguer: is it the creator? The viewer? Both? and if so, how do we model their `needs?' Additionally, good evaluations are expensive to implement and execute. While some communicative visualizations warrant this level of evaluation, many do not. We prefer a low-cost method to assess whether we are successful. Thus, the evaluation of communicative visualizations is uncommon. In a metastudy of 800 papers, those in the category of ``evaluating Communication through Visualization (CTV)'' were rarely found--four times in one study~\cite{lam:hal-00723057}--and none in another~\cite{isenbergctv}!

While we don't find many examples of actual evaluation, we do see advice on what form this evaluation might take. Questions and metrics include, ``(1) Do people learn better and/or faster using the visualization tool? (2) Is the tool helpful in explaining and communicating concepts to third parties? (3) How do people interact with visualizations installed in public areas? Are they used and/or useful? (4) Can useful information be extracted from a casual information visualization?''~\cite{lam:hal-00723057} At this level of abstraction, \textit{learning} is operationalized using distant metrics like `time-on-site' or `engagement' (e.g., clicks). The result is a big gap between concept and metric. With a more concrete language--and learning objectives in particular--we can have tighter integration between goals and assessments.

\subsection{Visualization as Teaching}

Those who produce visualizations often recognize that they have a mission to `educate' their viewers/readers. In a recent blog post, Jonathan Schwabish writes, ``...we all need to find ways to help our readers know what's important and what we want them to \textit{learn}'' (emphasis ours)~\cite{Schwabish}. We believe that many in the communicative visualization community have been circling around the idea that communicative visualizations and learning are connected.\footnote{An example quote: ``\ldots we aim to build some analysis into most of our graphics, arming readers with tools \textit{to understand} why races were won and lost and offering context.''~\cite{nytimeselec}} Without explicitly defining what it is that we want to teach, we have begun crafting strategies for better teaching! Techniques such as personalization~\cite{adar2017persalog}, explorable explanations~\cite{victor}, storytelling and narrative~\cite{Segel2010,Conlen2018}, active discussions around visualizations~\cite{Hullman2015com}, visual difficulties~\cite{hullman2011vd}, rhetorical strategies~\cite{rhetoric}, gamification~\cite{diakopoulos2011design}, and draw-your-own~\cite{nyt:draw,Kim2017} style interaction are all ways of improving learning and all are evident in the traditional learning sciences community. Being specific in how we define objectives would allow us to find the \textit{best} strategy for any problem~\cite{ambrose2010learning} and to better evaluate new approaches.

\subsection{Visualization Design Education}
Visualization instructors may recognize the generic student design specification: ``I want the user to understand insights in the data'' or worse, ``I want the user to explore the data.'' Work in visualization pedagogy has produced rubrics and techniques for evaluating student visualization work, but this often follows conventional lines: does the visualization express a set of facts and can it be effectively read~\cite{asee_peer_28150, Forsell:2010:HSE:1842993.1843029, Hearst:2016:EIV:2858036.2858280}. Neither students nor teachers have a clear way of articulating a design goal.  Instructors will resort to heuristic evaluation techniques which emphasize generic guidelines focusing on cognitive efficiency. Expert review~\cite{Tory:2005:EVE:1092228.1092239} is another option, but is likely too expensive for general use. Even expert critique would benefit from a formal specification of a designer's intent. More critically, assessing if a student has learned to design visualizations may be a poor proxy for assessing if the student has learned to design given intent. A clear objective language can greatly benefit the pedagogical practice (in fact, we use it in our classes).

In the context of visual analytics (both education and practice), there is an identified need for end-to-end evaluation~\cite{Munzner:2009:NMV:1638611.1639181}. Given particular risks in design choices (e.g., wrong encoding technique for data or missing interaction mechanisms), different `upstream' and `downstream' evaluations may mitigate those risks (e.g., interviews, lab experiments, etc.). Results from these evaluations are constantly fed back to the designer to improve the design~\cite{Sedlmair:2012:DSM:2720013.2720398}.  Within the learning sciences, we see a similar feedback loop. Because the experience of the student may be far removed from what the teacher intended~\cite{Lowe1988,hinds1999curse}, assessment helps close the loop~\cite{popham2003test}.   While many visualization systems papers focus on specific end-users (e.g., the biologists who might use the gene sequence visualization), recent work has shown that broader, crowd-based evaluations can be applied~\cite{Heer:2010:CGP:1753326.1753357}. With clear learning objectives and associated assessments, one could similarly test communicative visualizations in a crowd setting. 

\subsection{Learning Sciences}
One area that has influenced our thinking is the education literature on the use of graphics \textit{as part of} learning~\cite{clark2010graphics}. This line of work evaluated graphics (photographs, visualizations, maps, diagrams, etc.) in the context of a curriculum (e.g., ~\cite{7192676}) and demonstrated ways in which graphics can both enhance and detract from learning~\cite{clark2010graphics}.

From the assessment literature we draw on the idea that our primary goal for assessment is `programmatic.' We want to know that we designed our visualization (i.e., `the educational program') well \textit{across all readers}, and not that any \textit{specific reader} achieved our objective.

\begin{figure*}[htbp!]
  \centering
\includegraphics[width=\textwidth]{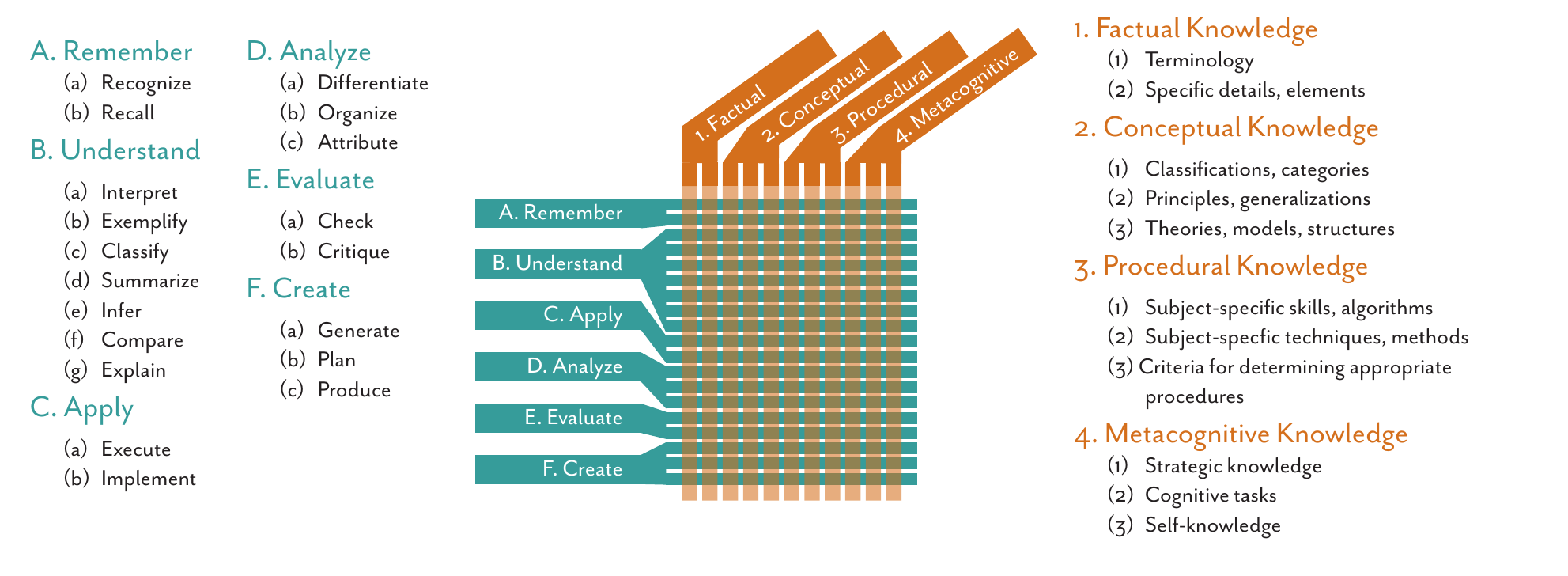}
  \caption{Cognitive Taxonomy, adapted from~\protect\cite{anderson2001taxonomy}. Learning objectives are constructed by selecting a row and column to identify the verb and noun for the objective: \textbf{The viewer will \textit{[verb]} \textit{[noun]}}. While verbs (cognitive constructs on the $y$--axis) can be used directly, the nouns (knowledge dimension on the $x$--) are category types that are replaced with specifics.}
  \label{fig:cogtax}
\end{figure*}

\subsection{Visual Literacy}
Visual literacy represents another bridge between communicative visualization and learning. Recent efforts in understanding, measuring, and improving visual literacy~\cite{alpers17,borner19,6875906,7539634} build on significant work from psychology (e.g., `graph sense'~\cite{friel01}). It is undeniable that visual literacy will be a mediator (among many others) in the effectiveness of a communicative visualization given a learning objective. In some cases, the designer may also have a metacognitive objective on improving improving visual literacy. We mean this in both the broad sense (learn how to read a bar chart) and in the specific (learn how to read my novel encoding that I want to use again next week). Our framing supports such objectives.

\subsection{Communication Theories}

Communication theories can be broadly split into \textit{process-focused} and \textit{meaning-focused}~\cite{fiske}. The former derives from Shannon and Weaver. The model centers on the mechanism by which a message is encoded by a sender and transmitted to a receiver~\cite{shannon}. Shannon and Weaver describe three levels of problems: How accurately can we transmit the message (level A)? How precisely does the transmission convey the meaning (level B)? How effectively does the received meaning affect conduct in the desired way (level C)? Level C is most related to our goals as it would, in theory, allow us to describe the deviation of what the sender wanted to affect (physically or cognitively) and what actually happened. Unfortunately, the classic process models focus on the lower level problems of accurate encoding in the presence of noisy channels. Nonetheless, the process model has impacted computer science broadly, and visualization specifically~\cite{chen10}.

The second theoretical line, the meaning-focused, is most commonly recognized as semiology or semiotics. The roles of different agents, signs, codes, signifiers, referents, etc. help model ``meaning making''~\cite{chandler}. Bertin's \textit{Semiology} is the clearest connection between this school and information visualization~\cite{bertin,Kindlmann,mackinlay1986automating}. Though he does not make explicit reference to any specific approach (i.e., Peirciean or Saussarian), Bertin offers a communication model focused on how meaning is formed through monosemic images. However, communicative visualizations (in contrast to archival or analytical) are not analyzed significantly. Bertin briefly offers that communicative visualizations should, ``create a memorizable image which inscribes THE OVERALL INFORMATION within the field of assimilated knowledge'' (note, \textit{memorizable} not \textit{memorable}). He expands: ``School maps, blackboard sketches, and all representations of a pedagogic nature to inscribe information in the viewer's memory, to make it become assimilated knowledge, capable of being \textit{recalled} at the time of an exam, a conversation, a research project, or a decision.'' While this teases at the relation of the designer to an instructor, Bertin largely stops there. Thus the underlying theory does not offer us a way to comprehensively model a designer's intent or evaluate the design's success.
\section{The Cognitive Domain}
To leverage existing learning objective taxonomies we considered a number of existing schemas including: Structure of Observed Learning Outcomes (SOLO)~\cite{biggs2014evaluating}, Understanding by Design (UbD)~\cite{wiggins2005understanding}, Data--Information--Knowledge--Wisdom (DIKW)~\cite{dikw}, and the `Revised' Bloom~\cite{anderson2001taxonomy}. Ultimately, we opted for the Revised Bloom as it was best developed and could clearly separate the idea that there were things people should be able to do (`verbs') as distinct from those things they should learn (`nouns').

Figure~\ref{fig:cogtax} illustrates the two dimensional model (cognitive processes $\times$ knowledge). The way to form a learning objective statement is to select a verb from the rows and a noun from the columns to fill in a sentence of the form: ``the viewer will [verb] [noun].'' 
Any particular visualization can have multiple objectives associated with it. For example, Figure~\ref{fig:phillips} shows how we build three different objectives for a visualization from the Economist~\cite{econcurve}. Note that while the verbs are generic, the nouns are replaced with specifics. Here, the visualization is communicating a change in the Phillips curve (inflation and unemployment have a stable and inverse relationship) over three time periods. The designer has a particular insight (the weakening correlation) that they would like the viewer to remember. The simplest objective, ``the viewer will \textit{recognize} the \textit{weakening correlation},'' reflects a simple intent related to this insight.  The slightly more complex objective, ``The viewer will \textit{classify} \textit{different temporal regimes for the Phillips curve},'' reflects a more complex insight, one with conceptual structure, that requires a higher level understanding. The appeal of this approach, is that we are often able to hold one facet (the noun or verb) constant while adjusting the other. For example, we can `upgrade' our verb to indicate that ``the viewer will be able to \textit{generate hypotheses} on why we have \textit{different temporal regimes}.''\footnote{Whether the visualization actually supports this objective is another matter--but one that can be assessed.} 

To provide examples below, we have used the narrative visualizations collected by Segel and Heer~\cite{Segel2010} and Hullman and Diakopoulos~\cite{rhetoric}. For each, we have `reverse engineered' plausible objectives from the text. This allows us to provide examples and also validate that the taxonomy covers the likely intents. We use both larger visual products (i.e., an entire narrative Web page or application with many visualizations and views) as well as individual visualizations. Where necessary, we collected additional examples. The taxnonomy and examples are available on our supplemental website.

\subsection{Cognitive Processes}
Within the Revised Bloom there are six main cognitive process categories with additional sub-categories (Figure~\ref{fig:cogtax} and Table~\ref{tab:cog} illustrate both categories and specific verb instances). The processes were originally intended to be hierarchical, with more complex or difficult processes (e.g., \textit{create}) encapsulating easier ones (e.g., \textit{recall}).\footnote{Whether this is a true hierarchy is the subject of much debate~\cite{haladyna2012developing} but is a debate we ignore for now.}

The base level of the taxonomy is \textit{remember} (with sub-types \textit{recognize} and \textit{recall}). This should be most familiar to visualization practitioners. It corresponds to the few studies of memory that do exist in the communicative visualization literature~\cite{doi:10.1177/107769908906600315,borkin16,hullman2011vd,Bateman:2010:UJE:1753326.1753716,83476,saket} as well as Bertin's model of a memorizable image.

In some ways, this dimension is also the easiest to design for. If the designer wants a viewer to remember that `low unemployment was associated with high inflation' (as they do in Figure~\ref{fig:phillips}), they can employ many visual tricks to call their attention to this point (annotation, different colors, etc.). This is not to say that simple insights are the only thing that is worth recalling. The designer may also want to ensure that critical definitions (e.g., the definition of `elbow point' as in Figure~\ref{fig:elbow}A) are learned. For example, the silhouette plot in Figure~\ref{fig:elbow}B actually came from a page describing the algorithm for using the plot~\cite{sil}. The designer's objective may be for ``\textit{the viewer to recall the procedure for using a silhouette plot}.''

\begin{figure*}[htbp!]
  \centering
\includegraphics[width=.9\textwidth]{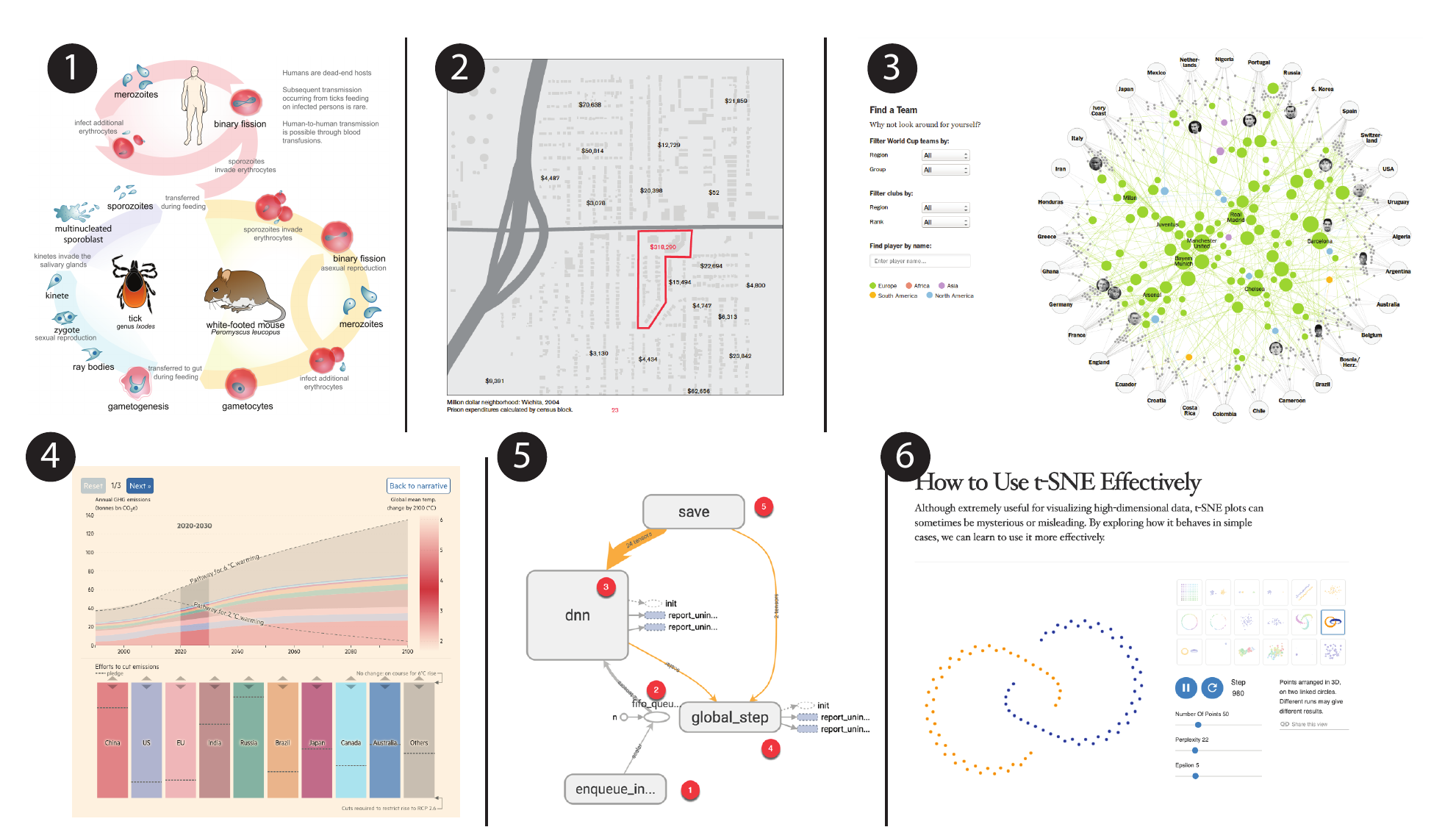}
  \caption{Example of communicative visualizations: (1) The Babesia life cycle in humans\cite{wiki:babs}, (2) Prison expenditures by census block from~\cite{crimeblocks}, (3) The final image from the `scrollytelling' vis: ``The Clubs That Connect the World Cup''~\cite{soccer}, (4) Interactive climate change calculator~\cite{changecalc}, (5) Annotated Tensorboard example~\cite{tensorboard}, and (6) An interactive lesson on ``how to use t-SNE effectively''~\cite{wattenberg2016how}}.
\label{fig:examples}
\end{figure*}

More broad than \textit{recall} is \textit{understand}--a term not so helpful on its own but one that encapsulates a number of useful verbs. Achieving learning of this type requires being able to summarize or explain certain phenomena or insights. An example would be a visualization of features of the Iran nuclear program~\cite{iran} with an associated objective of, ``\textit{summarizing the key components of the program}.'' With a network diagram of the lifecycle of the Babesia parasite~\cite{wiki:babs} an objective maybe be to: ``\textit{paraphrase the key stages of the parasite's development}'' (Figure~\ref{fig:examples}(1)).

Many \textit{explorable explanations}~\cite{victor} often fall into the \textit{apply} category. Distill's t-SNE tutorial, for example, offers to teach, ``how to use t-SNE effectively'' with a series of interactive visualizations~\cite{wattenberg2016how} (Figure~\ref{fig:examples}(6)). Here, the designer may intend the viewer, ``\textit{use the appropriate t-SNE settings for a given distribution}'' or ``\textit{implement the t-SNE algorithm}.''

\textit{Analysis} requires the integration of different pieces into a whole. This may mean contrasting smaller insights as with the different indicators on a thematic map tracking crime (population, socio-economics statistics, prisoner migration, etc.)~\cite{crimeblocks} (Figure~\ref{fig:examples}(2)). The designer may want the viewer to, ``\textit{contrast the indicators to identify disproportionate impact}.'' Fulfilling the objective requires the viewer pull specific data insights, understand their relation, and integrate them.

The \textit{evaluate} dimension is one that demands an ability to critique. In many cases, visualizations are built to support an argument. An analyst who had a hypothesis, and validated it, may summarize their result in a communicative visualization. The communicative form not only shows the insight but often provides evidence for it. John Snow's famous Ghost Map~\cite{snow1855mode} is a prime example where the goal was for the viewer to, ``\textit{critique the hypothesis that cholera was airborne}'' and ``\textit{judge as accurate the model of cholera as waterborne}.''

The final category, \textit{create} focuses on advanced cognitive processes such as being able to generate new theories. This objective puts the viewer in the role of a (passive) analyst. For example, the New York Times' visualized soccer team connections, showing the viewer interesting insights and encouraging them to \textit{generate} their own hypotheses~\cite{soccer} (Figure~\ref{fig:examples}(3)). This example also illustrates the interplay of narrative and learning objectives. The visualization follows a martini glass structure~\cite{Segel2010}, first conveying key insights to recall (the stem), and then encouraging a generative process (the glass).

\begin{table}[]
\begin{tabular}{p{2.2cm}p{6cm}}

\hline
\textbf{Category} & \textbf{Verbs} \\
\hline
\rowcolor{lgray}
A. Remember   & (a) recognize, identify;  (b) recall, retrieve                                                                                                                                                                                                                                          \\ 
B. Understand & (a) interpret, clarify, paraphrase, represent, translate; (b) exemplify, illustrate, instantiate; (c) classify, categorize, subsume; (d) summarize, abstract, generalize; (e) infer, conclude, extrapolate, interpolate, predict; (f) compare, contrast, map, match; (g) explain, model \\
\rowcolor{lgray}
C. Apply      & (a) execute, carry out; (b) implement, use                                                                                                                                                                                                                                              \\
D. Analyze    & (a) differentiate, discriminate, distinguish, focus, select; (b) organize, find,integrate, outline, parse, construct; (c) attribute, deconstruct                                                                                                                                        \\
\rowcolor{lgray}
E. Evaluate   & (a) check, coordinate, detect, monitor, test; (b) critique, judge                                                                                                                                                                                                                       \\
F. Create     & (a) generate, hypothesize; (b) plan, design; (c) produce, construct                                                                                                                                                                                                                     \\ \hline
\end{tabular}
\caption{Cognitive Dimension Verbs}
\label{tab:cog}
\end{table}

\subsection{Knowledge dimensions}
The currency of most information visualization is \textit{insight}. While there are various forms of insight~\cite{4797511}, the majority focus has been on insights derived through visual analysis by an analyst (e.g.,~\cite{Saraiya:2005:IME:1070610.1070747}). There are two problems using this model for communicative visualization. First, we must recall that there are two agents involved in communication--the designer and viewer. The designer makes the choice of whether something is an insight, whether it will be an insight to the viewer, and then decide if, and how, to communicate it in the design. Second, insights do not neatly capture everything we may want to communicate. For example, a \textit{process} or \textit{algorithm} may be something we would like to visually communicate. The algorithm may be an insight (i.e., a particular flow chart might describe it, and the viewer might remember it). However, if the objective is for the viewer to be able to \textit{use} the algorithm, `insight' does not sufficiently describe our goal.

Just as `recall' is a good start for our verbs, `insight' is a good layer on which to build our nouns. In fact, `viewer will recall an insight,' forms the basis of many communicative visualizations we experience. However, this form does not capture the real range of possibilities. The Revised Bloom knowledge dimensions provide an alternative starting. At the first level sits Factual Knowledge, which in many ways represents an atomic, specific insights. From this, the taxonomy builds upwards to include other types of knowledge.

At the simplest, \textit{Factual Knowledge}, includes the simplest facts and figures but also simple definitions. The insight that the optimum cluster size is 4 (Figure~\ref{fig:elbow}) may fall in this category. However, even this simple fact can be integrated into a more complex cognitive objective (e.g., ``to recall the number,'' ``to use the value to predict,'' etc.). In the visualization context it is likely that factual knowledge can be captured directly in the encoding of the visual mark (e.g., the number of cars sold in the height of a bar). Very important facts or insights--those the designer wants the viewer to remember--may be highlighted with annotations or alternative encodings.

\textit{Conceptual knowledge} requires the integration of multiple facts. For example, the Financial Times climate change calculator~\cite{changecalc} intends for the viewer to ``\textit{analyze the impact of different policy `bundles' on global temperature}'' (Figure~\ref{fig:examples}(4)).

\textit{Procedural knowledge} is most often related to skills or algorithms. Learning a procedure for visually identifying outliers in a scatter plot fits here (note, we want the viewer to identify outliers in \textit{any} scatter plot, not just in the single example they are given). A visualization of a deep neural net's structure in Google's TensorBoard~\cite{tensorboard} can help train a viewer to ``\textit{create a new workflow applied to their own data}'' (Figure~\ref{fig:examples}(5)). Flowcharts used for diagnosis or triage are in this category (medical, repair, etc.). Medical practitioners, for example, are taught to make decisions based on patient symptoms and other situational factors (e.g.,~\cite{esi12})--if a patient displays symptom $X$ do $Y$, otherwise check for symptom $Z$, etc. Flowchart style visualizations can support the learning of these `algorithms.'

Finally, \textit{metacognitive knowledge} relates to a viewer's knowledge about learning strategies, cognitive tasks, or self-knowledge. It would be rare for a \textit{single} communicative visualization to deliver this kind of knowledge on its own. However, a diet of visualizations that force the viewer to draw their own guesses may help the viewer, ``\textit{identify preconceptions and strategies to debias their thinking}''~\cite{Kim2017}. Visualizations that communicate uncertainty~\cite{SPEIER20061115,natter2005effects} may intend that the viewer, ``\textit{develop better/broader ways of understanding risks or making decisions}''. In educational contexts, instructors often employ metacognitive checklists or other visualizations (e.g.,~\cite{Robertsvisme}) that enable students to track their progress over time. These `quantified self' style visualizations are often generated by the learner themselves. They can not only track simple metrics like test performance, but more nuanced questions about the depth of understanding.

Any combination of verbs and nouns are plausible. However, some combinations may be more rare. For example, we have found far more recall $\times$ factual than create $\times$ metacognitive. This was done largely through our inference of intent. To better determine the true distribution and assess the usefulness of the learning objectives framework we describe a survey and interview study of designers.

\section{Survey Study}
While we are able to map many existing visualizations to our taxonomy, this was done through inference of intent. Ideally, we would like to understand if \textit{actual} design intents also fit. To test this, we recruited professional designers from the Data Visualization Society (DVS) to participate in a survey and interview study.

\subsection{Data Collection}
We reached out to people in the DVS Slack who posted their own visualization in either the \#share-critique or \#share-showcase channels. The visualizations in these channels were still being designed, or were recently finished, and the intent would likely be `fresh' in the minds of the designers. We sent individuals 3-question \textit{personalized} surveys, with the option of indicating interest in an interview. Of the 34 people who received the survey, 29 responded.

For each visualization-participant pair, we created customized learning objectives for the visualization and integrated these into the survey. These objectives reflected our best inference on the intents of the designer and utilized both the visualization itself and any information provided in the Slack post. Participants could also add their own objectives (and 23 added at least one). We provided an initial set of objectives to both reduce the participation burden and to model how objectives could be written. While this design allows for rapid data collection without training, it may create some default bias in responses. However, from those respondents whom we later interviewed, we do not believe this to be a concern. Future work may involve a more specific intervention to teach designers how to use the framework. Surveys were sent to the designer through a private message on Slack. On average, we created $\sim$4 objective statements (a minimum of 2 and maximum of 6--132 in total). Figure~\ref{fig:heatmap} (left) depicts the distribution of learning objectives \textit{we} proposed in the survey. 

Participants were asked to select those objectives that reflected their design intent, and optionally add learning objectives. Participants were not trained in the language of the taxonomy and could only model them on our examples. We manually mapped the participant provided statements to the taxonomy.  In most cases the mapping was clear. When the statement was not a cognitive learning objective we did not use it for this analysis (15 were removed). Figure~\ref{fig:blackgirlmagic} is an example image from one of our participants with their selected objectives.

\subsection{Data Analysis}
\begin{figure*}[htbp!]
\centering
\begin{minipage}[b]{\dimexpr.5\textwidth-1em}
  \centering
 \includegraphics[width=.9\linewidth]{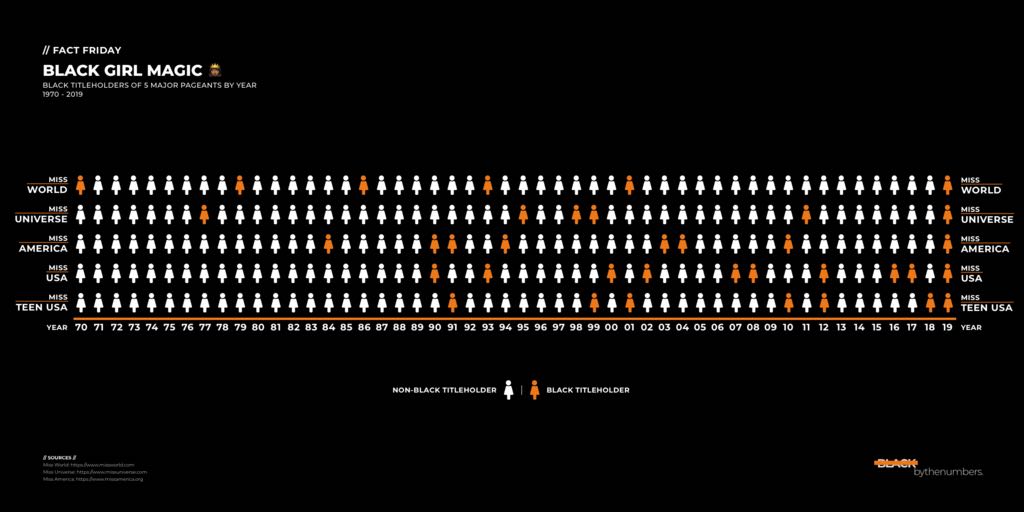}
  \caption{An example of a visualization from the DVS Slack Channel. The participant selected these cognitive learning objectives as goals: 1) Recall that all 5 major pageants have black titleholders in 2019. 2) Compare the ratio of non-black titleholders to black titleholders over the years. Visualization used with permission from participant, created by Black by the Numbers (www.blackbythenumbers.com).}
 \label{fig:blackgirlmagic}
\end{minipage}\hfill
\begin{minipage}[b]{\dimexpr.5\textwidth-1em}
  \centering
  \includegraphics[width=.9\linewidth]{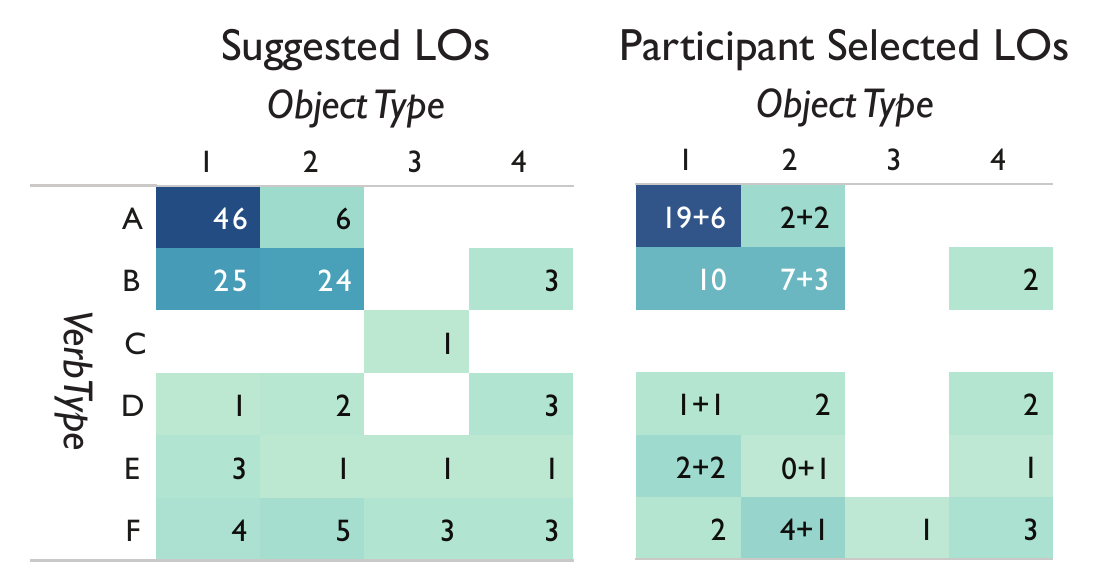}
  \caption{Distribution of suggested learning objectives sent in the surveys (left) and participant selected learning objectives from the survey responses (right). For the right panel we also indicate the number of new objectives generated by respondents (e.g., $+6$). Cells are color coded based on the fraction of objectives in that cell. The relative distributions between our inference and the participant responses were very similar.}
  \label{fig:heatmap}
\end{minipage}
\end{figure*}

Twenty-six out of 29 survey respondents reported that at least one of our learning objectives was something that they hoped their audience would be able to do after viewing their visualization. Figure~\ref{fig:heatmap} (right) reflects the distribution of the cognitive objectives selected or added by the designers. Participant-provided objectives ranged from other cognitive learning objectives (``Recall the high variability of [the variable] over time.'' [P3]), to affective goals (``Feel an emotional response to the issue'' [P4]), to other non-learning objectives, such as business goals (``Return to our website / subscribe to our newsletter'' [P17]). The three respondents that did not choose any cognitive learning objective mentioned an affective learning objective they were hoping to achieve (e.g., ``Feel an emotional response to the issue'' [P4]).  Though not the focus of this work, we consider these other intents in our discussion. 

As expected, most of the suggested and accepted learning objectives were at the lowest level of learning objectives with the verb ``Recall'' and the knowledge descriptor of ``Fact.'' An example objective from our survey is: ``The viewer will recall that there is no correlation between production budget and Oscar nominations.'' It is possible that because of our limited perspective on higher level design goals--we could see very little of the design process in the Slack channel--we focused on simpler objectives. However, the participants largely confirmed that this distribution was accurate both in their selections and in the interviews. Thus we have some confidence that it is possible to infer intents.

While participants did not agree with all our inferences, our interviews revealed some nuance to this. In some cases, participant selected the \textit{best} reflection of their intent and ignored the other plausible, but less ideal, choices. In a few cases, our choice may have surprised the designer. They were things the designer either hadn't considered or explicitly \textit{did not} want the audience to learn. One participant wanted to accurately communicate that the first category had a much higher value than the others, but not have the category overwhelm the others. ``The [category] is doing more than all my studying, more than all the others combined, which is interesting. But I wanted to be able to communicate that, I didn't want it just totally overwhelm." [P4] 
This response may reflect a need for \textit{anti-objectives}--an idea we return to later.

\section{Interviews}
To expand on the insights gained from the survey, we collected additional information through interviews.

\subsection{Data Collection}
Of the 29 survey respondents, 16 agreed to a 30 minute semi-structured video interview (a set of our starting questions is provided as supplementary information to this paper). Questions ranged from basic demographics (e.g. ``How long have you been making data visualizations?''), to more open-ended prompts (e.g. ``Can you tell me about your design process?''). Participants who were interviewed were compensated with a \$15 Amazon gift card. We asked a set of preliminary questions on the participants' background and experience with data visualizations. The interview largely focused on the visualization that participants had posted to the DVS Slack channel and their design process, design considerations, and audience. When possible, we broadened the focus to broader practice. Finally, we asked about their opinions on using learning objectives for visualizations and whether or not they could be applied to the participant's design process. 

\subsection{Data Analysis}
We transcribed our interviews and used open coding from a grounded theory approach.  In the initial pass, transcript segments were coded based on emerging themes. Codes were expanded over the initial coding pass. In a second iteration, interviews were re-coded based on the entire code set (130 unique codes). The most common codes were consolidated into four broader themes (process, decisions, goals, and design considerations). Below, we highlight key insights illuminating how designers conceive of goals, the potential usefulness and usability of learning objectives, and the relationship to non-cognitive objectives.

\subsection{Results}
When reflecting on their design process, several participants (6) were explicit in that \textit{goals} or \textit{take-aways} were an important starting point for them. One participant describes his process as: ``\textit{I'm always thinking, `What do I want to show? And what story lies behind behind the data?' So that's my thinking, always the first point of thinking about visualization.}'' [P2] He imagined the goal of his visualization as communication of a story, which was a common theme among our participants. P1 noted, ``\textit{You have to start with the premise for which you want someone to take away from this and then say, `How do I depict that to accomplish that goal?'}''  Only two participants explicitly indicated that they had, ``\textit{never thought about that actually.}'' [P3] 

\subsubsection{Impact on Design}
Of our 16 interview participants, 15 responded that learning objectives would be helpful or useful. A telling example was one participant who said, ``\textit{when I design a chart, I want somebody to look at it and be able to understand it quickly. I think, I don’t remember what the chart says a few days later.}' [P10] But reflecting on the learning objectives said:

\begin{displayquote}
\textit{It didn't really occur to me that, ``Oh, I really want this, I want people to think about this or I want people to recall this and apply it to their own lives.'' I found that having to articulate that and think about that, I found that very helpful. So yeah, learning objective, that's very useful. I think in my day to day life we talk a lot about users, and a lot of the times it's sort of providing information. But I don't think about information that lasts, right?} [P10] 
\end{displayquote}

The last point validates our view that design often focuses on cognitive efficiency rather than lasting impact. Learning objectives could potentially help designers think more about their intent for what they want a viewer to remember, understand, evaluate, or take action on for such impact.

In the case of this last participant, introducing the idea of learning objectives appeared to have an immediate effect on design:

\begin{displayquote}
\textit{\ldots [in] your survey where you were like ``what do you want people to take away from this?'' I hadn’t really thought about it. The other thing I wanted people to take away was [this insight] so I was like, ``okay, I’ll make that more prominent this time around.''} [P10]
\end{displayquote}

While this provides some anecdotal evidence that designers can leverage learning objectives in their process, a more formal evaluation would be useful. Put another way, the question is: do designers make \textit{different} choices when using learning objectives?

\subsubsection{Learning to Use Learning Objectives}
While the objectives taxonomy provides a formal way of modeling intent, it is a complex tool to use. In our interviews, we often found it difficult to `teach' participants deeply about learning objectives. One participant noted that, ``\textit{I think it's a useful way to think about it, but it's hard as well.}'' [P4] This is in line with the education community that has found that the process of creating learning objectives requires training~\cite{gronlund2008gronlund} and may be difficult even with practice~\cite{raible2016writing}. One limitation of our study, and the use of learning objectives in general, is that our participants may need significant training to fully understand learning objectives and their use. 

We also found that experience with teaching or teachers helped participants understand the function and usefulness of the objectives. Five of our participants indicated they had experience teaching or had someone close to them who is a teacher. All five thought that using learning objectives could be useful or helpful to them. When asked about their opinion of using learning objectives for visualizations, one participant says ``\textit{I never thought about that actually, but I think it would work very well, I mean, I'm, for example, sometimes I'm doing lectures here at University and they're I think exactly like that.}'' [P2]

\subsubsection{Limits of Cognitive Learning Objectives}
While we have focused on \textit{cognitive} learning objectives, there are other valid objectives a designer might have. Both our surveys and interviews emphasize that \textit{cognitive} learning objectives are not the only goal a designer might have. For example, many of our participants noted affective goals, and three participants \textit{only} had affective goals. These affective goals all focused on evoking an emotional response from the audience, such as to perceive ``\textit{the tragedy}'' [P13], ``\textit{appreciate the diversity}'' [P15], ``\textit{feel an emotional response to the issue}'' [P4], and ``\textit{being proud.}'' [P2]

An open question is the relationship of the taxonomy we have created to analytical visualization tasks. Two participants in our study reflected this concern.
``\textit{I would look into the [learning objectives] which I'm answering when I'm trying to communicate a story out of it. In that sense, yes. But when I'm just exploring the data set, I'm not so sure.}'' [P5] 
Similarly, one participant noted that there is an exploratory phase of looking at the data, where goals may shift.
Because of this, sometimes learning objectives may not be clear to the designer at the beginning or may change during the design process. 
``\textit{So I would say learning objectives are good place of course to start, but that definitely learning objectives change after it's been built or after you've taken a closer look at the data.}'' [P6]

Analytical and communicative visualizations are not necessarily on a continuum but can work hand-in-hand. Just as we may implement `analyst' features in a communicative visualization (querying, insight building, etc.) we may have communicative goals embedded in exploratory software. Future work may help understand how the objectives framework can be embedded into the design of visualization tools for analysts.

\section{Discussion, Limitations, and Future Work}
The use of the learning objectives framework requires a shift in thinking. We reflect on how learning objectives can be learned, extended, and used, as well as their limitations.

\subsection{Visualization Pedagogy}
We have been experimenting with learning objectives in an educational context. In our graduate classes we have a communicative visualization project that requires students to articulate both learning objectives (selecting from our modified cognitive taxonomy) and assessments.  Our belief is that for students, using learning objectives may allow them to (a) direct design iterations, (b) more formally compare alternatives, (c) have confidence that their objectives have been met, and (d) reduce post-hoc rationalization of their design choices. 

More research is needed to evaluate this approach. However, we have anecdotally found that students can better formulate their intents when guided through creating learning objectives. For example, one team took on visualizing forest fire data. They began with an under-specified proposal: ``Our objective is to teach viewers some factors of forest fires.'' With feedback, the team was able to break this into more specific learning objectives including: ``Viewers can describe which kind of tree is more likely to cause a fire,'' and, ``Viewers should be able to describe the periodical pattern of forest fire in a year.'' This change was due to acknowledging specific motivating insights: that factors included seasonality and tree types (some trees, with specific features, more likely to be involved in fires). 

Using the learning objectives framework will likely never make design more efficient or fast. However, we believe that it will make communicative visualizations better. With practice, building objectives and assessments is a process that can be internalized by a designer. In many cases, the designer may be able to mentally `simulate' the execution of a full-blown assessment and achieve the same result as actually running it.

\subsection{Affective, Business, and Archival Objectives}

As our interviews revealed, not all objectives fit in the cognitive framework. It is worth considering which frameworks may be needed to supplement this taxonomy.

\textbf{Affective Objectives}---A recent debate over whether visualizations can or should produce an empathetic response highlight that the cognitive-boost only formulation is not universally held. Speaking on empathy, Alberto Cairo stated, ``I am just very skeptical to the idea that data visualization is a medium that can convey (or even care about conveying) or increase `empathy'''~\cite{ctweet}. As a response, the artist Steve Lambert (and others) pointed to the case of the ``Gun Deaths'' visualization produced by Periscopic~\cite{guns}. Among other features, the visualization animates a tally of `stolen years' due to gun death. ``The tone is solemn,'' Lambert writes~\cite{lambert}, ``[the visualization] has its own pace, it conveys death, points toward the lost potential, and backs everything up with facts. It is an unqualified success in getting the participant's attention and interest, communicating the issue and data behind it. It works emotionally, leaving one troubled.'' He continues: ``Once one sets a clear objective, to reduce gun deaths in this example, learning this reality is revealed as just one of the early steps. \ldots \textit{good data visualization can expose people to the issue, capture their attention and interest, and even spur them to act for change}'' (emphasis ours). The quote illustrates both the clear cognitive objective--insights about gun deaths--but also an objective centered around a set of values. This is where the exclusive focus on the cognitive domain fails us.  Though such \textit{affective} objectives are not the focus of this work, we refer the interested reader to our supplemental site which offers a modification of an affective taxonomy~\cite{krathwohl1956taxonomy} as a potential starting point.

A structured formulation of affective objectives allows a designer, if they so choose, to specify that the viewer \textit{should} change their belief. Anti-objectives may be particularly useful in expressing neutrality through the affective taxonomy (i.e., ``viewer will \textit{not} modify their opinion''). In this sense, neutrality is a value that can be encoded as a learning objective. Perhaps more importantly, a principled language will support principled critique. This may support evaluation of topics such as deception~\cite{Pandey:2015:DDV:2702123.2702608} or `black-hat' visualizations~\cite{correll2017black}.

\textbf{Business Objectives}---One of the `goals' we often see articulated by content creators is that visualizations should increase engagement. A reasonable question is how a designer should represent engagement in the taxonomy framework? We argue that they shouldn't. One reason to desire increased engagement is to increase profit~\footnote{``Once I heard someone state: The purpose of visualization is funding, not insight.''~\cite{wijkvov}}. This is a perfectly acceptable \textit{business objective} but we argue that it is best measured through more direct means--like time spent on site or conversion rate. A second reason for increasing engagement is that it represents an instructional \textit{strategy} (i.e., more time studying $=$ more learning). If this is true, we suggest that the designer be direct about the \textit{actual} objective (recall, affect change, etc.), rather than proxy measures (time-on-site).

\textbf{Archival Objectives}---The third of Bertin's purposes for visualization, after analysis and communication, is ``Recording Information'' (or ``Inventory Drawings''). Visualizations that serve this function exclusively are ideally suited for the cognitive efficiency argument. Information from these types of plots should be extracted reliably. However, we argue that visualizations that exclusively have this function are rare. A paper may display results to allow others to find key data, but such visualizations often have a communicative intent as well (e.g., to draw attention to some key points). As with purely cognitive objectives, multiple high level objectives (e.g., business and communicative; communicative and archival) may require trade-offs. Future work may identify how to best navigate these.

\subsection{Multiple Objectives and Anti-Objectives}
Designing visualizations, even communicative ones, is a wicked design problem~\cite{buchanan1992wicked}. Multiple objectives and limited resources (e.g., the space on the screen or reader attention) means that a single visualization may not be able to solve all objectives equally, or at all. Crafting objectives for sophisticated visualizations also requires prioritizing them and using assessments to determine if those priorities are met. For example, we may be willing to sacrifice some performance on the \textit{recall} objective if the readers perform well in \textit{critique}.

One benefit of mapping multiple objectives onto the hierarchical matrix structure (e.g., Figure~\ref{fig:phillips}) is that a designer may be able to identify dependencies in objectives. The hierarchical nature of the framework allows one to order objectives. High level verbs (e.g., evaluate) and knowledge categories (e.g., procedural) either require expertise or indicate that scaffolding is necessary in the visualization. When ordered this way, a designer may be able to identify dependencies--to do \textit{Learning Objective 2}, my reader will first need to learn to do \textit{Learning Objective 1}. Future work may focus on the relationship of learning objectives to each other as well as the interplay between objectives and expertise (i.e., a difficult procedural task for a novice may be a simple recall one for an expert).

Interestingly, the learning objectives framework also allow us to indicate which things we don't want. This relates to Mackinlay's expressiveness test which emphasized that only those facts that we want to express were apparent in the visualization (and no others)~\cite{mackinlay1986automating}. We can be explicit about undesirable objectives, or \textit{anti-objectives}. An anti-objective takes the form of, ``\textit{a viewer will \textit{not} \ldots}'' A successful design ensures that objectives are met and the anti-objectives are not.

\subsection{Visualization Contexts}
A key observation that emerges from our construction of learning objectives and from our interviews is that the right level of detail for objectives requires considering the embedding context of the visualization. Objectives can be defined at many layers: for the single visualization, for a bundle of text and images, or even for an entire website.\footnote{We're fans of Tufte's view that, ``Evidence that bears on questions of any complexity typically involves multiple forms of discourse.''~\cite{tufte2006beautiful}} When we define learning objectives, we can't ignore the context in which the visualization is embedded. However, if we wish to evaluate a communicative visualization in context, it is useful to define a learning objective that can reasonably be influenced by that visualization. That is, the viewer should perform \textit{better} having seen the visualization than without it. More analysis of varying contexts is a crucial next step. An area of future work may also be to understand how the learning objective framework can be integrated directly with tools built for communicative visualizations (e.g., Tableau Stories, Idyll~\cite{Conlen2018}, and litvis~\cite{wood2018design}).

\subsection{Assessment}
The learning objectives framework is attractive, in part, because objectives are intended to be testable~\cite{popham2003test,   wiggins2005understanding}. By applying a formal language, our objectives are consistent with the `SMART' framework--Specific, Measurable, Attainable, Realistic, and Time Bound). Assessment protocols can readily be used to measure whether a viewer has achieved our learning objective.  We leave the design of specific assessments to future work. However, we briefly provide a few higher level observations and some examples.

As a simple example, given the Phillips curve chart (Figure~\ref{fig:phillips}), to assess the objective \textit{the viewer will classify different temporal regimes for the curve} we could ask: ``In which of the following periods did correlation vanish between unemployment and inflation (label each as true-false): (a) 1980-1990? (b) 1991-2000? (c) 2001-2010?'' If the viewer could not answer this question correctly \textit{before} they saw the visualization, but could \textit{after}, we could argue that we achieved the objective. Higher-level objectives (e.g., \textit{critique} or \textit{explain}) might use more open-ended responses. For example, for the objective \textit{the viewer will be able to explain why the Phillips curve has been flatlining in the past decade} we may need a grade rubric (e.g., full scores for describing all factors).

The appeal of assessments of this type is that we can more completely understand whether viewers \textit{attain} an objective given their initial state. One could offer a test before the viewer sees the visualization, one during exposure, and one after we take the visualization away. Separating the questions this way lets us explicitly determine if the visualization helped us achieve our intent: of the people who couldn't do or didn't believe something before, how many can/do after? 

The goal of assessment is to help the designer understand the impact of their design. This is distinct from the main goal of traditional assessment: to gauge the cognitive development of an individual student. We want to understand the performance of the intervention--the `program effects' or `lesson effectiveness.' That said, we leave open-ended what `program' means. While our goal is to help design single visualizations, we recognize that the communicative `intervention' can also include the combination of visualization and text, multiple visualizations, scientific papers, videos, or even entire websites. 

It is also worth acknowledging that while assessments \textit{can} be implemented (e.g., we can run an experiment on a crowdsourced platform), a designer may also benefit from merely \textit{articulating} what the assessment would look like. We anticipate that building and internally validating the assessment protocol (in relation to the objectives) would help the designer. At the very least, the designer can determine if the visualization conveys enough information for the objective to be achievable (e.g., the viewer won't be able to tell us about the 1995 Phillips curve if the data isn't present).

It is an open research question which assessment format will work best for communicative visualization learning objectives specifically. Assessments fall broadly into two categories: \textit{constructed response} (CR)--or open-ended style questions; and \textit{multiple-choice} (MC) formats. Each format has its own advantages and disadvantages. When both are suitable, MC forms are attractive as they are familiar, have been demonstrated to be effective, have been validated~\cite{haladyna2012developing} and are readily adaptable to computer-based testing~\cite{scalise2006computer}. Even though we may not yet know how to best assess communicative visualization learning objectives, the literature provides us with a good starting point.

\vspace{-3mm}
\section{Conclusion}
Designers of communicative visualizations almost always focus on the fidelity of the message. However, the viewer's ability to correctly \textit{decode} the message is at best a poor proxy for measuring if the message had the intended impact. In this work, we put forward the idea that we can formally model designer goals in a learning context. By framing communicative objectives as learning objectives, we allow a designer to describe their intent (using a variant of the Revised Bloom grid). Because such well-formed objectives can be readily transformed to assessments, this approach means that communicative designs can be better evaluated. By surveying and interviewing design professionals, we have demonstrated that intents can be mapped to learning objectives but that not all objective types are equally used in communicative visualization contexts. Our framing lends itself to future work in helping designers create objectives and transform these to assessment instruments. While existing visualization task and evaluation taxonomies largely focus on analytical/exploratory tasks, we offer learning objectives as a way of filling the gap for communicative visualizations.
\subsection*{Acknowledgements}

We would like to thank members of the DVS for their participation. We would also like to thank Jessica Hullman, Licia He, and Hari Subramonyam for their feedback, and John Stasko and Alberto Cairo for their comments on earlier versions of this work. We are grateful to the NSF for their support of this work through NSF IIS-1815760.

\balance{}

\bibliographystyle{abbrv-doi}
\bibliography{refs}

\end{document}